# EPICS FOR PDAs

Kenneth Evans, Jr., ANL, Argonne, IL 60439, USA

Abstract

With the advent of readily available wireless communications and small hand-held computers, commonly known as personal digital assistants (PDAs), it is interesting to consider using these portable devices to access a control system. We have successfully ported the Experimental Physics and Industrial Control System (EPICS) to Windows CE 3.0 (WCE) for the Pocket PC (PPC), and this paper describes the issues involved. The PPC was chosen because the WCE application programming interface (API) for the PPC is a subset of the Win32 API, which EPICS already supports, and because PPC devices tend to have more memory than other PDAs. PDAs provide several ways to connect to a network, using wired or wireless Compact Flash or PCMCIA Ethernet cards and modems. It is the recent advent of readily available wireless networks that makes using the portable PDA interesting. The status and issues surrounding the various kinds of wireless systems available are presented.

## 1 INTRODUCTION

WCE is sufficiently powerful to run EPICS, and it is sufficiently like Win32 to make porting it a feasible thing to do. It is sufficiently different from Win32, however, to make the porting nontrivial. Not all of EPICS was ported — only the part, Channel Access (CA), that is necessary to run client applications. The port was done by making projects in Microsoft eMbedded Visual C++ 3 (EVC), copying the existing EPICS files to those projects, and modifying the files inside the projects  No attempt was made to use the EPICS build system. The three necessary projects are dynamic link libraries for CA and the EPICS common library, libCom, plus the required executable, caRepeater. Parts of libCom that were not needed were omitted if problems arose. The version of EPICS Base used was 3.13.

One of the reasons for using EVC is that PPC devices use a variety of processors, and the executables need to be made for each processor. EVC is set up to do this, and the EPICS 3.13 build system is not. Moreover, this was a prototype endeavor, and it was not deemed feasible to make extensive changes in code that must be used for already-supported platforms. The disadvantage of this approach is that ongoing modifications to EPICS Base do not get automatically incorporated in the project files. The intent is for this port to be a prototype and to do the port again, based on the experience gained, and to use EPICS Base 3.14, which has better support for multiple processors.

Once the EPICS base modules are available, it is still necessary to write an application in order to use them and see if it all works. This has been done, and screen shots of two of the applications, ProbeCE and the APS Beam Display, are shown in Figs. 1 and 2.

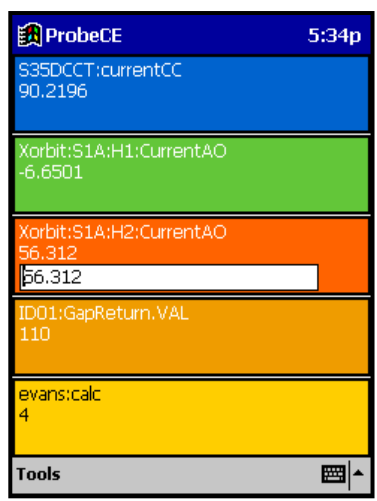

Figure 1: ProbeCE, a Windows CE application for EPICS. You can enter a process variable in any of the areas to monitor it. You can also change its value, as has been done in the middle area. There are additional features, not shown.

## 2 PROBLEMS

The major problems in porting CA to WCE involved, as might be expected, features in Win32 that are not supported in WCE.

### 2.1 Environment

One of the most significant unsupported features is that WCE does not have an environment and environment variables, as do most other common operating systems. Consequently, API functions such as getenv() do not exist. To get around this,  the environment variables that CA uses are stored in the registry and functions such as getenv() were rewritten

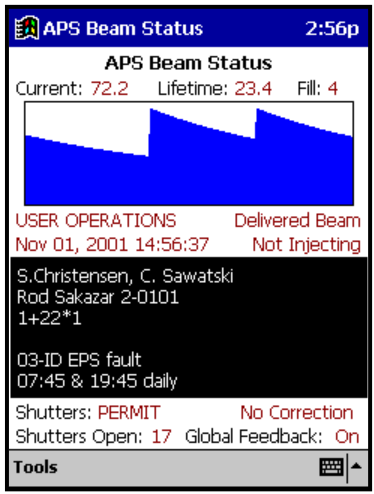

Figure 2: APS Beam Display, a Windows CE application for EPICS. This application shows the beam history and other relevant information. It updates when the information changes.

to access them from the registry instead of from the environment.

## 2.2 User Name

There is no user name as it exists in other common operating systems, and Win32 functions such as GetUserName() are not supported. To get around this, the user name, which is an essential part of CA, is implemented in the registry in the same way as environment variables. The consequence of this is that the user name can (and must) be set in the same way that environment variables can be set. In fact, the ProbeCE interface allows the user to set them all in the same dialog box, as shown in Fig. 3.

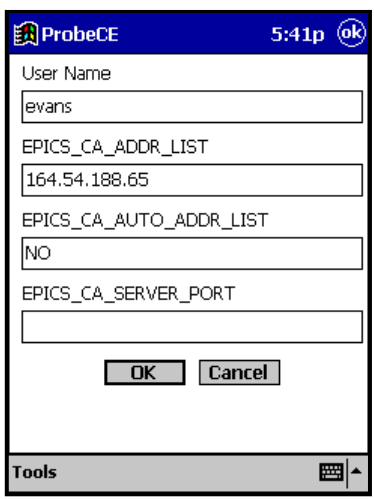

Figure 3: A dialog box for entering environment variables needed by EPICS. The user name is also entered here and is treated the same as an environment variable, possibly a security problem. WCE, however, has no authenticated user name.

Since the user can set whatever user name he desires, no matter how it is implemented, he can potentially get access to items in the control system that he was not intended to have just by changing the user name to that of a user that does have access. There is no authentication of the user name by a password as on more secure systems, such as UNIX and Windows NT/2000/XP. The security issue, however, is no more of a problem than with Windows 95/98/ME, where the user can log in with essentially any name he chooses. It is just more obvious, since the user is required to specify a name in order to use CA, as in Fig. 3.

The bottom line is that security must be enforced by some other means than by user name alone. EPICS Access Security has the ability to do this, though it may be inconvenient and use processor time.

## 2.3 Other

There were also problems with other common features that are not supported by WCE. These include many relatively standard UNIX-based functions such as exit(), abort(), many of the time functions, and functions related to errno, which does not exist. WCE only supports Winsock 1.1, whereas CA uses Winsock 2.0. There is no console in WCE and no support for routines such as printf(). These problems were worked around in various ways, which are probably of interest only to programmers, and there is insufficient room to describe them here.

## 3 WIRELESS

Network adapter cards, both wired and wireless are available for PPCs. The wired variety is more readily available, but typically where there is a socket into which to plug the wire, there is also a standard-size computer nearby. In such a case it is usually easier to use the larger computer with its bigger screen and more powerful operating system. The real incentive to use a PPC is that it is portable and easy to carry around. To maintain this advantage, it needs a wireless type network adapter.

Similarly, modems are available for the PPC. The most common type is one that is connected by a wire to a telephone outlet and can be used to access a dial-up network. It is also possible to connect to a cell phone in several ways, then use the cell phone to connect wirelessly. This has the disadvantage of having to manage two devices, the PDA and the cell phone, but allows portability. Finally, there are modem cards for the PPC that allow it to be used without cables or a cell phone.

Security is an issue for any type of wireless, since the data is transmitted by radio waves, which can be readily intercepted. Current technologies should probably not be considered secure in themselves, but the use of secure shells or Virtual Private Networks (VPNs) should adequately address the security problem at the expense of decreasing the data transmission rate. The new version of the PPC operating system, Pocket PC 2002, includes a VPN, and VPN clients are available for other VPN protocols.

Wireless networks can be loosely organized into three classes: wide area (WAN), local area (LAN) and short range or personal area (PAN). The following describes some of the popular current standards that are relevant to EPICS, but a complete discussion of the available technologies is beyond the scope of this paper.

*3.1 WAN*

Cell phones are the most common WAN wireless devices. Several of the available WAN protocols have data capability, and some can be used without requiring a cell phone. The advantage of using a WAN is that you can access the control system from far away, even another city. A disadvantage is that the WAN data protocols are usually slow compared to wired networks.

Cellular Digital Packet Data (CDPD) is one of the protocols for digital data transmission that does not require a cell phone. It does not, in fact, support voice at all. It uses a modem card rather than a network access card and has a maximum transmission rate of 19.2 Kbps. Several companies provide CDPD services for the PPC. CDPD is available in most large cities in the United States, but not elsewhere. CDPD can be used to connect to an Internet services provider (ISP) that can provide access to the control system. Security can be provided by using a VPN. The EPICS applications described above have been successfully used with CDPD along with a VPN for security.

Code Division Multiple Access (CDMA) is a protocol with a maximum transmission rate of 14.4 Kbps. It supports voice and data.

Global System for Mobile Communication (GSM) is a slower, older protocol with transmission rates to 9.6 Kbps. It supports voice and data. General Packet Radio Service (GPRS) is an enhancement for GSM that is widely available in Europe and some United States cities. It has a maximum transmission rate of 171 Kbps.

CDMA and GSM/GPRS cell phones can be connected to a PDA to provide wireless connections to a controls network, and there are modem cards available, as described above for CDPD, that do not require a cell phone.

*3.2 LAN*

The LAN network range indoors is on the order of 100 m, the size of a small building. Outside, the range is 1000 m or more. The prevailing standard and primary candidate of interest is 802.11b. It has a maximum transmission rate of 11 Mbps, the fastest of the currently available technologies. Networks and network cards for PDAs are available. The principal problem with 802.11b is security. Using the current encryption schemes, a sophisticated eavesdropper can capture a small amount of traffic and determine the encryption key in less than a hour. However, security can be provided by using a VPN. The EPICS applications described above have been run successfully on an 802.11b network with and without a VPN.

The next generation, 802.11a, should provide speeds to 54 Mbps and better encryption. It also uses a frequency band that is less likely to interfere with other wireless devices including cell phones and X10 devices.

*3.3 PAN*

A PAN network has a range of about 10 m. The primary PAN technology is Bluetooth with transmission rates to 1 Mbps. The intended use is for vending machines, ticket machines, etc., and as a replacement for cables, for example, to connect to a printer. Many companies have committed to Bluetooth, but significant implementation is yet to come. It is probably not useful for a control system, at least for the types of applications considered here. However, eventually you may be able to connect to your cell phone without a wire and use that to connect to EPICS.

## 4 CONCLUSIONS

PPCs provide a convenient, portable means of accessing a control system. EPICS CA has been ported to PPCs and applications have been developed that demonstrate that using a PPC to access a control system is quite feasible. Portability requires wireless network adapters or modems, and there are several available for PPCs. The wireless field is still in an early and rapidly changing state, but several viable usable technologies exist. Security is a problem that needs to be carefully considered, but secure shells or VPNs should provide adequate security, at the expense of some loss in data transmission rates.

## 5 ACKNOWLEDGEMENTS

The author would like to acknowledge valuable discussions with Andrew Johnson, William P. McDowell, and Steven Shoaf. This work was supported by the U.S. Department of Energy, Office of Basic Energy Sciences under Contract No. W-31-109-ENG-38.